\documentclass[twocolumn,aps,floatfix]{revtex4}
\usepackage{amssymb} \usepackage{graphicx} \usepackage{amsmath}
\usepackage[T1]{fontenc} \usepackage{pstricks} \usepackage{subfigure}
\usepackage[colorlinks=true,citecolor=blue,urlcolor=red]{hyperref}
\usepackage[normalem]{ulem}

\usepackage{hyperref}

\begin{document}

\title{Bistability of Bose-Fermi mixtures }

\author{Tomasz Karpiuk,$\,^1$ Mariusz Gajda,$\,^2$ and Miros{\l}aw Brewczyk$\,^1$}

\affiliation{\mbox{$^1$ Wydzia{\l} Fizyki, Uniwersytet w Bia{\l}ymstoku,  ul. K. Cio{\l}kowskiego 1L, 15-245 Bia{\l}ystok, Poland} \\
\mbox{$^2$ Institute of Physics, Polish Academy of Sciences, Aleja Lotnik{\'o}w 32/46, PL-02668 Warsaw, Poland} \\
}

\date{\today}

\begin{abstract}
We study the properties of the Bose-Fermi mixture from the perspective of reaching a state of a self-bound quantum droplet. The variational analysis shows that the system exhibits bistability. For weak repulsion between bosons, one of the equilibrium states, smaller in size, spherically symmetric, and with negative energy, corresponds to quantum droplet, the other with always positive energy represents the elongated droplet-like state immersed in the sea of a fermionic cloud. For stronger repulsion between bosons the bifurcation is seized and only the former state is left.  Now it represents an elongated object which, for strong enough boson-fermion attraction, gets negative energy. It becomes an excited Bose-Fermi droplet when the trap is released, what is demonstrated by solving the quantum hydrodynamics equations for the Bose-Fermi system. To depict our ideas we consider the $^{133}$Cs-$^6$Li mixture under ideal conditions, i.e. we assume no losses.
\end{abstract}

\maketitle

Quantum mixtures of atomic gases have been studied already for years, both experimentally and theoretically. Of particular interest are degenerate Bose-Fermi mixtures. While initially experiments were aimed at the realization of a degenerate Fermi gas via sympathetic cooling, where the bosonic or fermionic component served as a coolant part \cite{Hulet01,Schreck01,Hadzibabic02,Roati02,Aubin06}, soon after studies of many-body quantum phenomena started.

The phase diagram of the harmonically trapped mixture of bosonic rubidium-$87$ and fermionic potassium-$40$ atoms was determined in \cite{Ospelkaus06,Zaccanti06}. Tuning the interspecies interactions by Feshbach resonance, the experimentalists observed a collapse of the mixture above some threshold and on the attractive side of the resonance as well as the phase separation on the repulsive side. Binary mixture of potassium-$41$ condensate immersed in a sea of fermionic lithium-$6$ atoms was recently realized \cite{Grimm18} and collective oscillations of bosonic component were studied. A breathing mode of a condensate was induced by increasing the strength of the boson-fermion forces. On the positive side of Feshbach resonance the phase separation is eventually reached and the oscillation frequencies are measured \cite{Grimm19}. Here, temperature effects seem to be crucial to explain the experimental results, as shown in \cite{Grochowski19}. Physics of polarons was probed via radio-frequency spectroscopy of a gas of potassium-$40$ fermionic impurities immersed in an ultracold atomic gas of rubidium-$87$ \cite{Cornell16}. The energy, spectral width, and the lifetime of Bose polaron were determined on both sides of a heteronuclear Feshbach resonance, going beyond the standard textbook description in a weakly interacting regime.

A novel quantum mixture, a self-bound system of ultracold atoms has been realized experimentally recently. First quantum droplets were observed in a gas of dysprosium-$164$ atoms, which possess the largest dipolar magnetic moment among atoms \cite{Schmitt16}. Interplay of short range and dipolar forces is crucial for stabilizing such systems. More dipolar droplets were created soon with erbium-$166$ atoms \cite{Chomaz16}. Demonstration of existence of different kind of self-bound objects, a two-component mixture of bosonic potassium-$39$ atoms has followed \cite{Cabrera17,Semeghini18}. The origin of self-confinement both in dipolar systems as well as in two-component mixtures is due to quantum fluctuations which play essential role at the border of a collapse, as suggested in Ref. \cite{Petrov15}. Quantum droplets in a heteronuclear bosonic mixtures were also observed \cite{Fort19}. New self-bound systems, the Bose-Fermi droplets have been recently elaborated theoretically \cite{Rakshit19a,Rakshit19b}.

In Bose-Fermi mixtures, interactions among degenerate identical fermions can be changed by a contact with a Bose-Einstein condensate. For example, a mixture of $^{87}$Rb\,-$^{40}$K was used to bring a gas of fermionic potassium to collapse \cite{Modugno02,Bongs06}. Indeed, large enough boson-fermion attraction in rubidium-potassium mixture results in an effective attraction between fermions, responsible for the collapse. This attraction, on the other hand, could lead to fermionic superfluidity as in the case of phonon-induced attraction between electrons in superconductors. In fact, a Bose-Fermi mixture in which both the fermionic and the bosonic components are superfluid has been produced \cite{Ferrier14,Delehaye15}.

In Bose-Fermi systems, fermions can mediate the interactions between bosons as well. This kind of behavior has been recently observed  in an experiment with a mixture of ultracold fermionic $^6$Li and bosonic $^{133}$Cs atoms \cite{Chin19}. If degenerate lithium and condensed cesium atoms attract each other strongly enough it may result in effective attractive boson-boson interactions even though the condensed bosons alone are repulsive. This, mediated by fermions, change of the interaction character is caused by a coherent three-body scattering process. It leads to formation of trains of Bose-Fermi solitons seen in experiment \cite{Chin19} and predicted theoretically a long time ago \cite{Karpiuk04,Santhanam06,Karpiuk06}.

A crude estimation of the onset of a change of the sign of interactions among bosons, given in Ref. \cite{Karpiuk04}, can be displayed as $g_B n_B = |g_{BF}| n_F$, where $n_B$ and $n_F$ are the densities of bosonic and fermionic fractions, respectively, taken at the center of the trap. The parameters $g_B$ and $g_{BF}$ determine the strength of contact interactions. The densities can be roughly estimated assuming that the density of each component is calculated within the Thomas-Fermi approximation while the other component is not present. Then for  sufficiently large Bose-Fermi attraction
\begin{equation}
\frac{|g_{BF}|}{g_B} > C\, \frac{N_B^{2/5}}{N_F^{1/2}} 
\label{cr_ratio}
\end{equation}
the bosonic component becomes effectively attractive too. Here, $C=C_1 (a_\perp^B/a_{B})^{3/5} (a_\perp^F/a_\perp^B)^3 \lambda_B^{2/5}/\lambda_F^{1/2}$, $C_1=3^{9/10} 5^{2/5} \pi /16$, $a_{\perp}$ is the radial harmonic oscillator length, $\lambda=\omega_z / \omega_\perp$ defines the aspect ratio of the axially symmetric (around $z$ axis) trap, and $a_{B}$ is the $s$-wave scattering length for the pure Bose gas related to the interaction strength via $g_B=4 \pi \hbar^2 a_{B}/m_B$. The interaction between bosons and fermions is characterized by $g_{BF}=2 \pi \hbar^2 a_{BF}/\mu$ with $\mu$ being the reduced mass.

To meet the condition (\ref{cr_ratio}) tuning one of scattering lengths via Feshbach resonance technique is required. It can be done in two ways: i) by tuning $a_B$ to very small values or ii) by tuning $|a_{BF}|$ to large values. 

\begin{figure}[bth] 
\includegraphics[width=6.5cm]{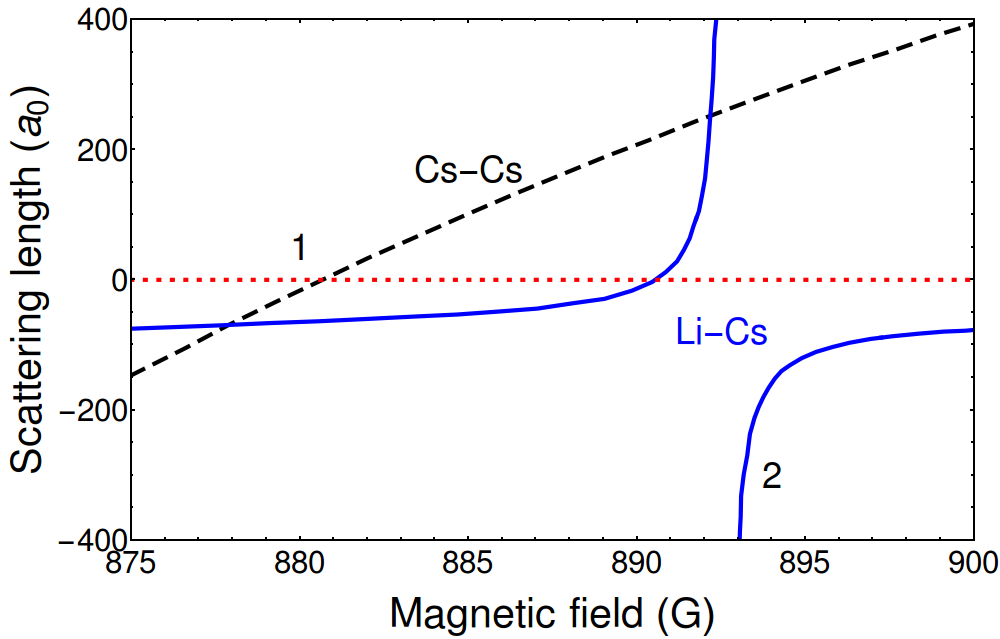}
\caption{Scattering lengths for Cs-Cs (dashed black line) and Li-Cs (solid blue line) collisions as functions of the magnetic field (see Ref. \cite{Chin19}).  } 
\label{LiCs}
\end{figure}

We consider small $a_B$ case first. For cesium-lithium mixture as in experiment of Ref. \cite{Chin19}, i.e. for $a_{BF}\,\approx -60\, a_0$ ($a_0$ is the Bohr radius), trapping frequencies $(130,130,6.5)\,$Hz for bosons and $(400,400,36)\,$Hz for fermions, and particles numbers as $N_B=30000$ and $N_F=20000$, the Eq. (\ref{cr_ratio}) gives the upper limit $a_{B} < 0.4\,a_0$ (region marked by $1$ in Fig. \ref{LiCs}). In fact, the soliton trains were observed in \cite{Chin19} already for $a_{B}\,\approx 3\,a_0$. 

The other way to satisfy the condition (\ref{cr_ratio}) is to work close to the Feshbach resonance visible in Fig. \ref{LiCs}, at the magnetic fields around $893\,$G. Across this resonance one has $a_{B}\,\approx 230\,a_0$ and the Eq. (\ref{cr_ratio}) is satisfied for large enough $|a_{BF}|$, equal about $2000\,a_0$ (region marked by $2$ in Fig. \ref{LiCs}). Negative and large values of $a_{BF}$ are explored in Ref. \cite{Chin17}, where the observation of a degenerate Fermi gas trapped by the Bose-Einstein condensate is reported.

In the following we will discuss a scenario of droplet formation. We consider a trapped case with fixed number of fermionic, $N_F$, and bosonic, $N_B$, atoms in a harmonic trap. Although a genuine droplet must be stable without any trapping potential, initially atoms are always kept in a trap. Only then interaction parameters are tuned to required values and a droplet is eventually formed. To this end both kinds of atoms must be mixed in the well prescribed proportions. Typically these condition is not met in a trap with initially prepared atoms.  However, droplet can still be formed but then a gaseous cloud of surplus atoms is surrounding a droplet. The cloud is stopped from expansion by the external trap. The liquid droplet is  at a stable equilibrium with its  vapors then.

This is the situation we are studying here.  We show that Bose-Fermi mixture, spanned over parameter regions $1$ and $2$ in Fig. \ref{LiCs}, shows a bistability phenomenon while trapped by external harmonic potential. For a given set of interaction parameters, there are two minima in the energy landscape, one local and  the other global one. Density profiles correspond to two different droplet-like objects immersed in a vapor of surrounding atoms.

We start first with the variational calculations. The bosonic and fermionic densities are assumed to be axially symmetric distributions of the form of the Gaussian functions with two variational parameters: the axial ($\sigma_a$) and radial ($\sigma_r$) widths. The width of the bosonic cloud is the same as that of the fermionic one. No fermionic background is assumed. These assumptions are simplistic, nevertheless quite well illustrate the main idea. Hence, the densities are proportional to $\exp{(-z^2/\sigma_a^2)} \exp{(-\rho^2/\sigma_r^2)}$ and normalized to the number of fermions $N_F$ and bosons $N_B$, respectively. The kinetic fermionic, including the Weizs\"acker contribution \cite{Weizsacker}, and kinetic bosonic energies are then given by
\begin{eqnarray}
E_{kin}^F  &\propto&     1/\sigma_a^{2/3} \sigma_r^{4/3}          \nonumber \\
E_W   &\propto&   \left( 1/\sigma_a^2 + 2/\sigma_r^2 \right)      \nonumber \\
E_{kin}^B  &\propto&  \left( 1/2\sigma_a^2 + 1/\sigma_r^2 \right)  \,.
\label{ekinetic}
\end{eqnarray}
At the mean-field level, the only interaction energies are $E_{int}^B\,, E_{int}^{BF} \propto 1/\sigma_a \sigma_r^2$ as degenerate fermions do not interact. We consider as well the quantum corrections to the intra- and inter-species interaction energies. The Lee-Huang-Yang correction \cite{Lee57} to the boson-boson interaction is calculated as $E_{LHY} \propto 1/\sigma_a^{3/2} \sigma_r^3$, while the quantum correction to the boson-fermion interaction \cite{Viverit02}, $E_q^{BF}$, is given in the Appendix \ref{energy}. Finally, the contributions due to trapping are as follows
\begin{eqnarray}
&&E_{tr}^{B}  \propto  ((\omega_a^B)^2\, \sigma_a^2 /2 + (\omega_r^B)^2\,  \sigma_r^2 )    \nonumber \\  
&&E_{tr}^{F}  \propto  ((\omega_a^F)^2\, \sigma_a^2 /2 + (\omega_r^F)^2\,  \sigma_r^2 )  \,,   
\label{etrap}
\end{eqnarray}
where $\omega_a^B$ ($\omega_a^F$) and $\omega_r^B$ ($\omega_r^F$) are axial and radial trapping frequencies, respectively, for bosons (fermions). Full formulas for all kinds of energy included in the analysis can be found in the Appendix \ref{energy}.

\begin{figure}[bth] 
\includegraphics[width=7.5cm]{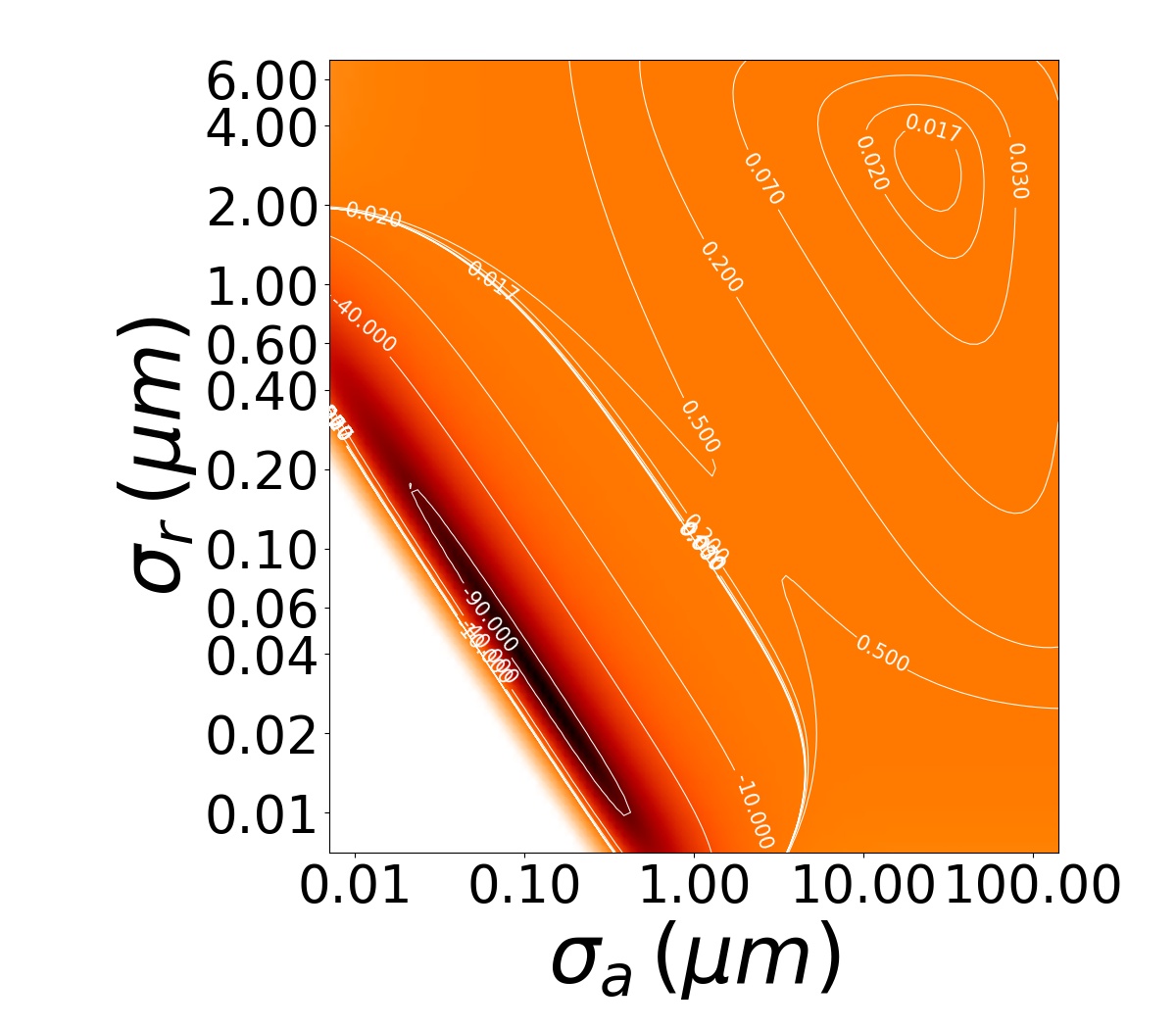} 
\caption{Equipotential lines for bosonic component of the Bose-Fermi mixture. Here, the trapping frequencies are $(130,130,6.5)\,$Hz for bosons and $(400,400,36)\,$Hz for fermions, $a_{B}=3\,a_0$, $a_{BF}=-60\,a_0$, and the number of particles are $N_B=30000$ and $N_F=3000$. The figure exhibits minima at the widths $(\sigma_a, \sigma_r)=(60,3)$ micrometers (with positive energy) and $(\sigma_a, \sigma_r) = (75,75)$ nanometers (with negative energy). For $N_F=2\times 10^4$ there is only the minimum at $(60,3)$ micrometers. } 
\label{varana1}
\end{figure}

The variational analysis described above confirms the bistability phenomenon, i.e. the existence of two minima in the energy landscape of a trapped Bose-Fermi mixture. The first minimum is found to describe the large in size and axially symmetric mixture, whereas at the second minimum the mixture forms rather a spherically symmetric droplet. We show the equipotential lines in Fig. \ref{varana1} (here, $a_{B}=3\,a_0$) for the set of parameters close to the region marked by $1$ in Fig. \ref{LiCs}, explored in experimental work of Ref. \cite{Chin19}. For $N_B=30000$ and small number of fermions (here, $N_F=3000$) we always find two energy minima. One of them, corresponding to a positive energy, represents the elongated axially symmetric object with the axial and radial density widths equal to $(\sigma_a,\sigma_r)=(60,3)$ micrometers. 

The second minimum corresponds to the spherically symmetric atomic clouds. For lower $a_{B}$ energy at this minimum  is negative, i.e. the object, in principle, should remain stable after the trap is removed since the trapping energy is positive. The size of the `droplet' is below $100\,$nm and increases with $a_{B}$. Its energy increases with $a_{B}$ as well and already for $a_{B}=20\,a_0$ becomes positive and larger than the energy corresponding to the first minimum. The atomic densities of spherically symmetric objects are large, too large to resist destruction due to three-body losses. Hence, the regime $1$ in Fig. \ref{LiCs} constitutes rather a good set of parameters for investigating trains of Bose-Fermi solitons. They appear as a result of the modulational instability of bosonic component of elongated object assigned to the first minimum (as already observed in \cite{Chin19}). For larger number of fermions, $N_F=20000$, there exists only one minimum with positive energy. This can be understood based on the Bose-Fermi droplet's physics -- a particular number of fermions can be associated with a given number of bosons only to form a droplet. The surplus particles, fermions in this case, act as a gas being at equilibrium with the droplet.

%\begin{figure}[bth] 
%\includegraphics[width=7.5cm]{energy_Gauss2_03_ab10_abf-60_00.jpg} 
%\caption{Equipotential lines for bosonic component of the Bose-Fermi mixture for $a_{B}=10\,a_0$, $a_{BF}=-60\,a_0$, and $N_B=30000$ and $N_F=3000$. Still two minima are present, one corresponding to negative, the other to positive energy. For $N_F=2\times 10^4$ there is only one minimum with positive energy. The barrier separating two minima is here about $10$ times higher than in the case of Fig. \ref{varana1}. } 
%\label{varana2}
%\end{figure}

%\begin{figure}[bth] 
%\includegraphics[width=7.5cm]{energy_Gauss2_03_ab20_abf-60_00.jpg} 
%\caption{Equipotential lines for bosonic component of the Bose-Fermi mixture for $a_{B}=20\,a_0$, $a_{BF}=-60\,a_0$, and $N_B=30000$ and $N_F=3000$. There are still two minima, one minimum at $(60,3)$ micrometers with positive energy and the second one at about $(250,250)$ nanometers but this time with positive energy much higher than the other minimum. } 
%\label{varana3}
%\end{figure}

\begin{figure}[bth] 
\includegraphics[width=7.7cm]{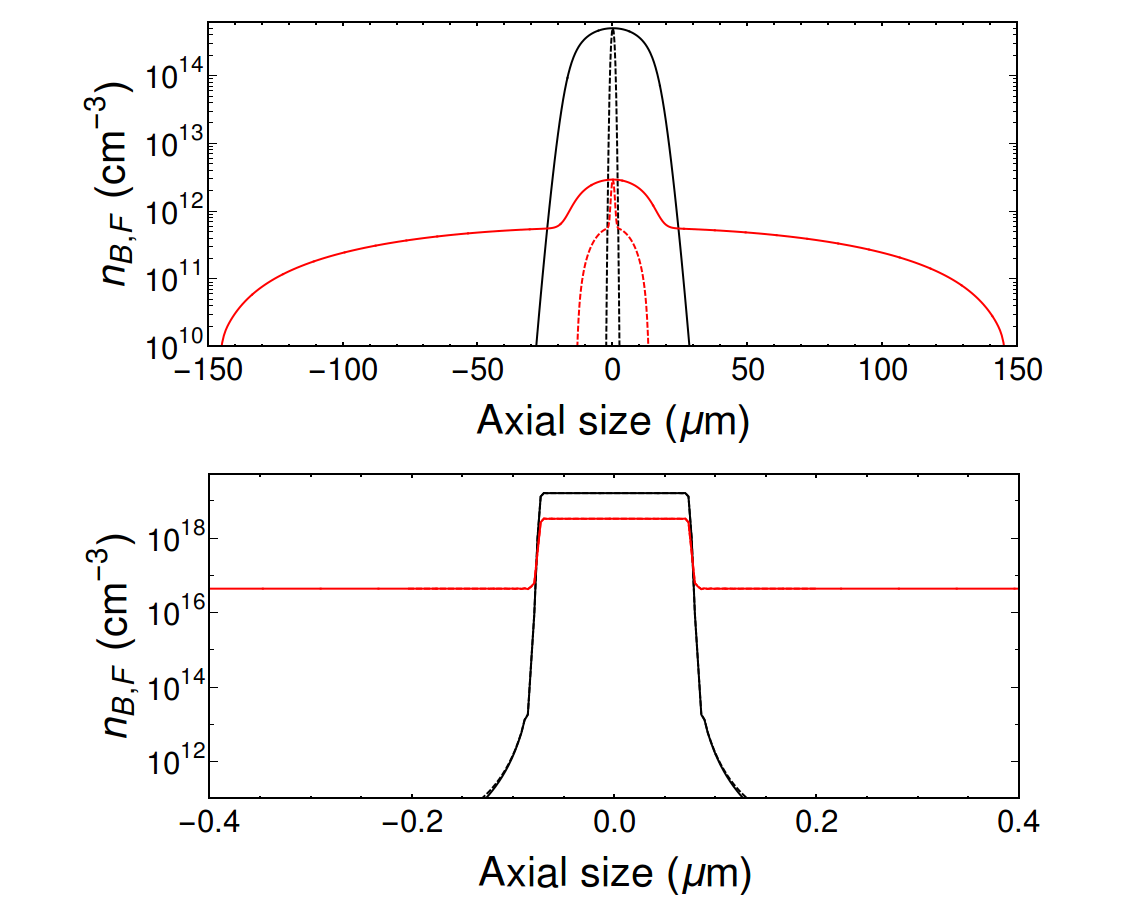} 
\caption{Ground state densities of the mixture obtained by using the imaginary time technique. For $N_B=3\times 10^4$, $N_F=2\times 10^4$, $a_{B}=3\,a_0$, and $a_{BF}=-60\,a_0$ we find a droplet at the energy minimum (lower frame). The figure shows bosonic (black, solid for axial and dashed for radial) and fermionic (red) densities. The droplet is spherically symmetric with the radius equal to about $100\,$nm. The small radius and high number of atoms result in very large densities. There is about $3000$ fermions in the droplet. Surplus fermions form a background, the level of which depends on the numerical box used in simulations. Lower densities can be reached for $a_{B}=10\,a_0$ and still $N_F=20000$ (upper frame). Now the bosonic densities are of the order of $10^{14}\,$cm$^{-3}$. An elongated droplet of size $(-30,30)$ in axial and $(-3,3)$ micrometers in radial directions (the ratio is about $10$), immersed in a fermionic sea, is clearly visible. } 
\label{imaginary}
\end{figure}

To verify predictions obtained within the variational analysis we performed numerical simulations based on quantum hydrodynamics equations \cite{Rakshit19a,Rakshit19b}. We look for the ground state of the system using imaginary time technique. For $a_{B}=3\,a_0$ and $a_{BF}=-60\,a_0$, while observing as consecutive states approach the ground state  corresponding to the droplet (lower frame in Fig. \ref{imaginary}), we see a long plateau in the energy plot as a function of imaginary time. We interpret this plateau as a property indicating the existence of a local minimum corresponding to an elongated object in Fig. \ref{varana1}. When the real time evolution starting with any density distribution belonging to this long plateau is done, we observe a quick (in duration less than $1\,$ms) change towards the state with the droplet, the surplus fermions are being expelled into the fermionic background. To avoid  falling on the second, stable, minimum one should increase $a_{B}$. Analogous simulations for $a_{B}=10\,a_0$ reveal a long-living droplet-like object (upper frame in Fig. \ref{imaginary}). This happens because, as shown by variational calculations, for $a_{B}=10\,a_0$ there exists a huge barrier separating the two minima. After removing trapping potential this droplet-like object breaks into a train of Bose-Fermi solitons.

\begin{figure}[bth] 
\includegraphics[width=7.7cm]{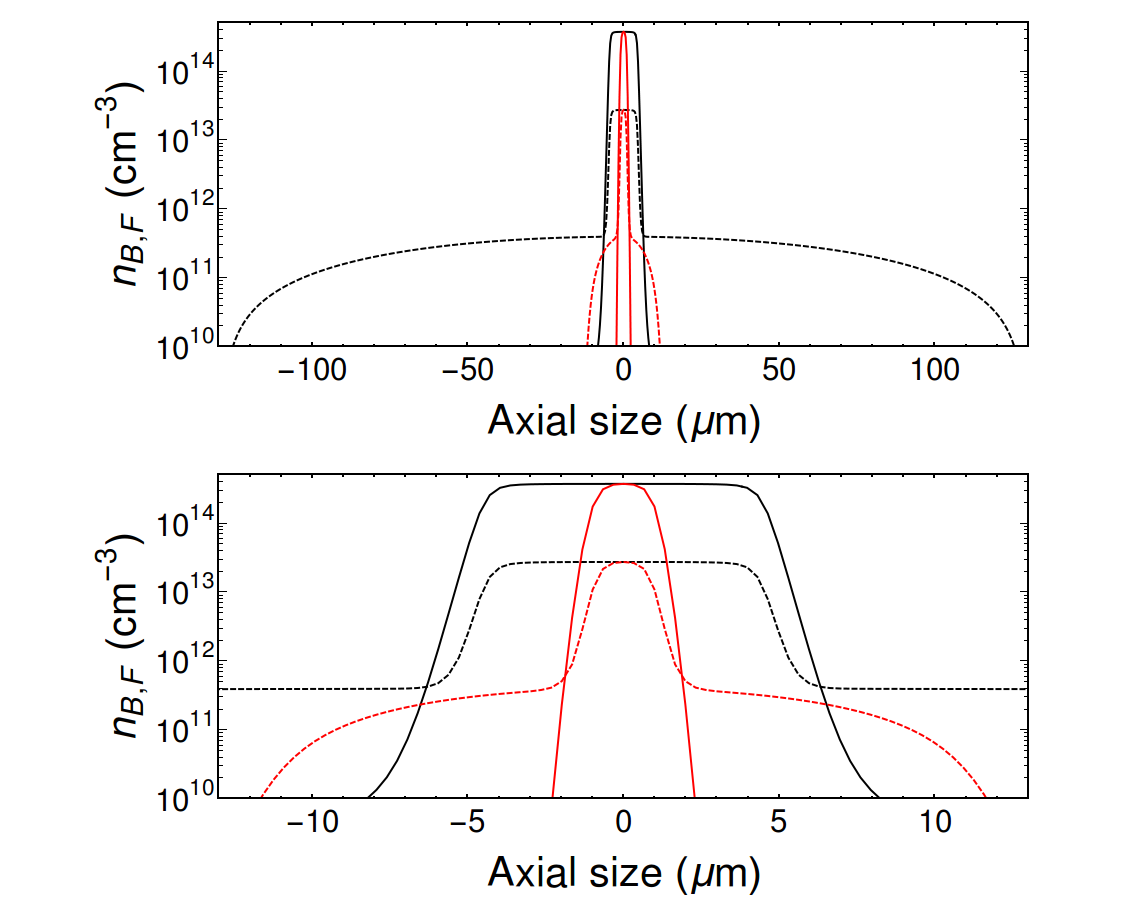} 
\caption{Ground state densities of the Bose-Fermi mixture for the case $2$ in Fig. \ref{LiCs}. Here, $a_{B}=250\,a_0$, $a_{BF}=-2.8\,a_{B}$, $N_B=10^4$, $N_F=10^3$, and the trapping frequencies are $(130,130,6.5)\,$Hz for bosons and $(400,400,36)\,$Hz for fermions. The figure shows bosonic (solid black and red lines) and fermionic (dashed lines) densities. The lower frame is just the zoom of the upper frame. } 
\label{imaginaryII}
\end{figure}

To get Bose-Fermi cesium-lithium droplets of larger size, to meet the imaging resolution requirements, one has to move from region $1$ to $2$ in Fig. \ref{LiCs}, i.e. to increase the value of $a_{B}$ scattering length. It simultaneously results in decrease of droplet component's densities making the particle losses less important. An example is shown in Fig. \ref{imaginaryII}, where the ground state densities, still at the presence of the trap, obtained by imaginary time evolution of quantum hydrodynamics equations for $a_{B}=250\,a_0$ and $a_{BF}=-2.8\,a_{B}$ are shown.

\begin{figure}[bth] 
\includegraphics[width=6.0cm]{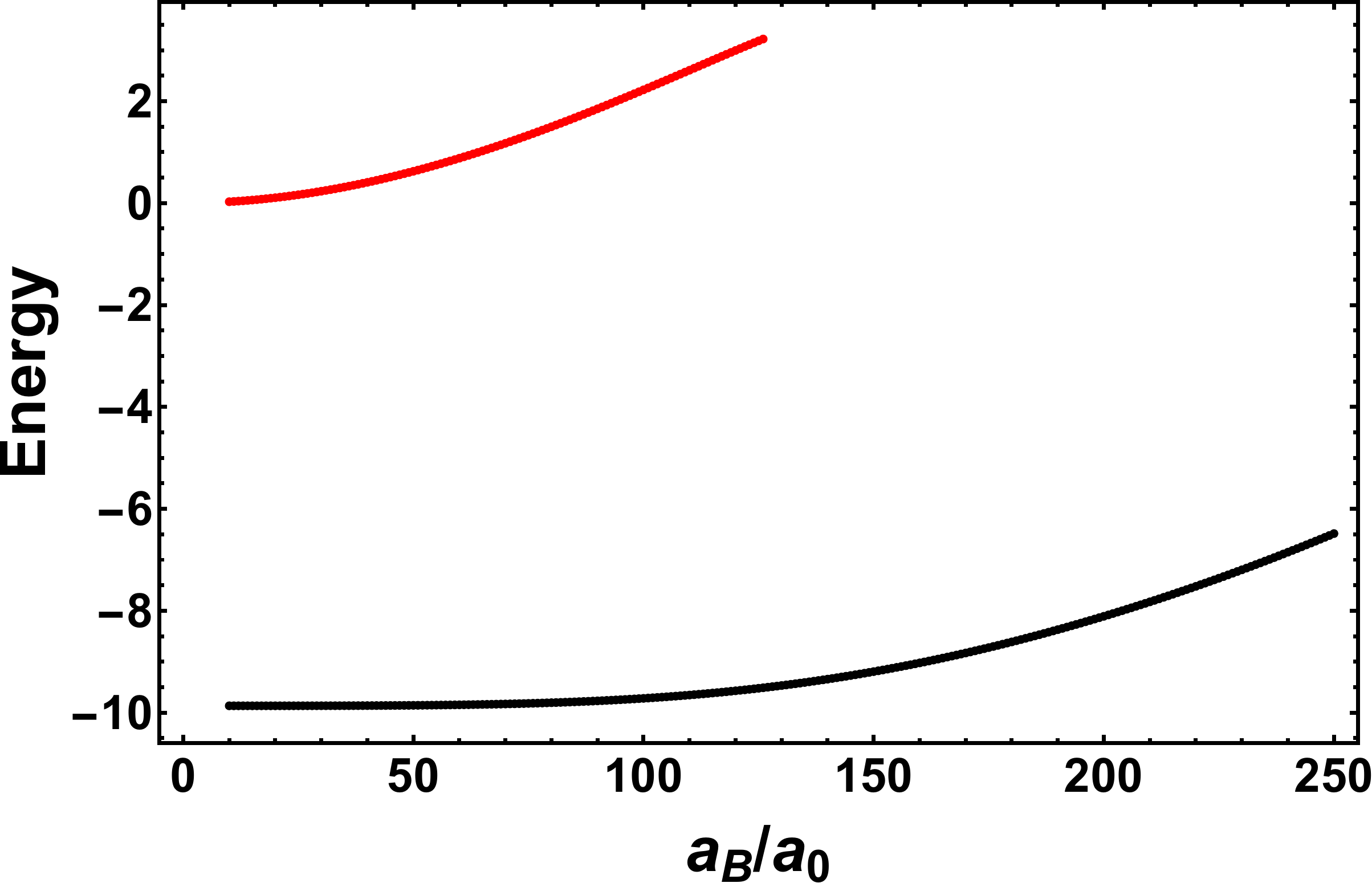} \\
\includegraphics[width=6.0cm]{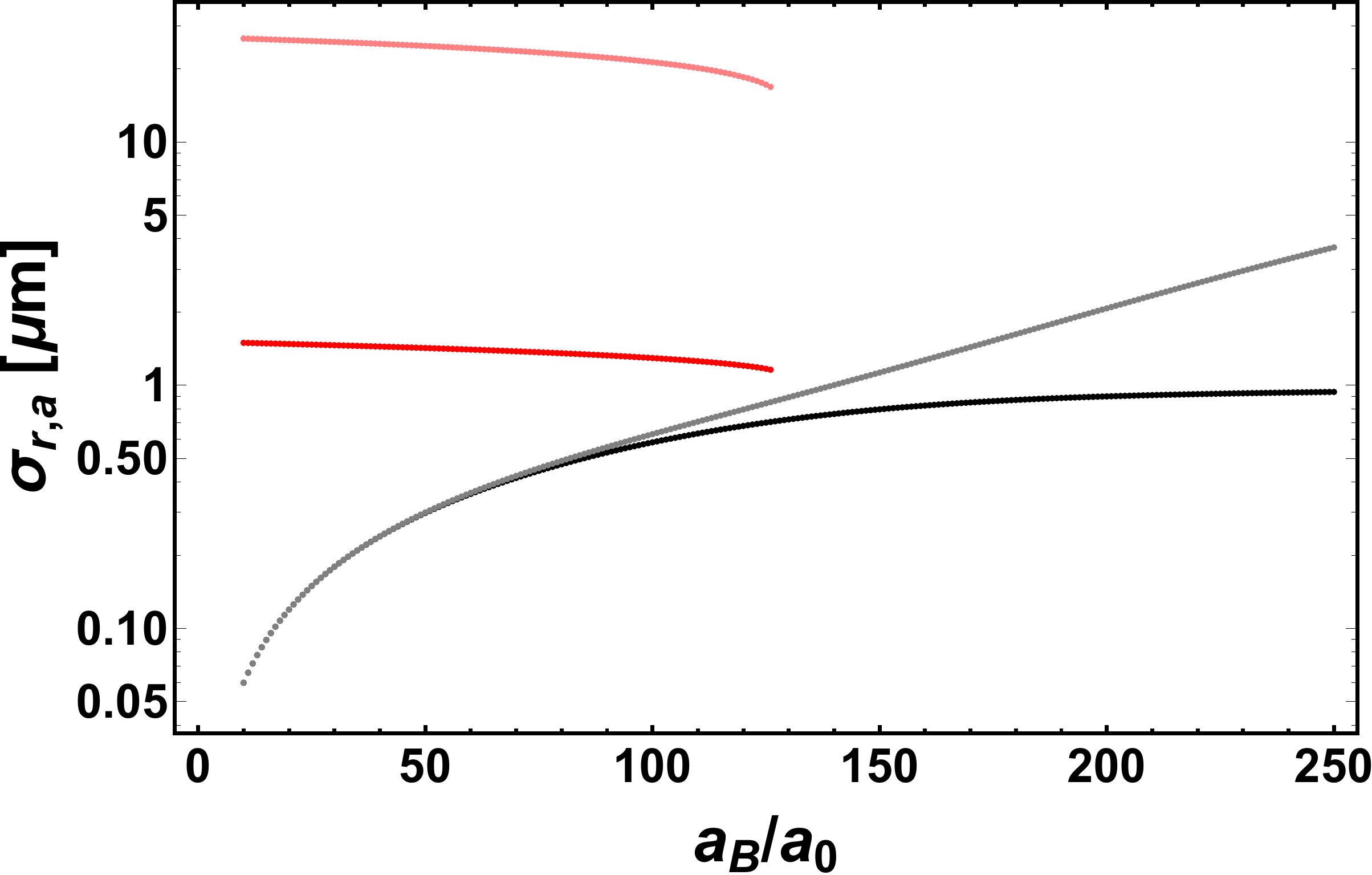}   \\
\includegraphics[width=6.0cm]{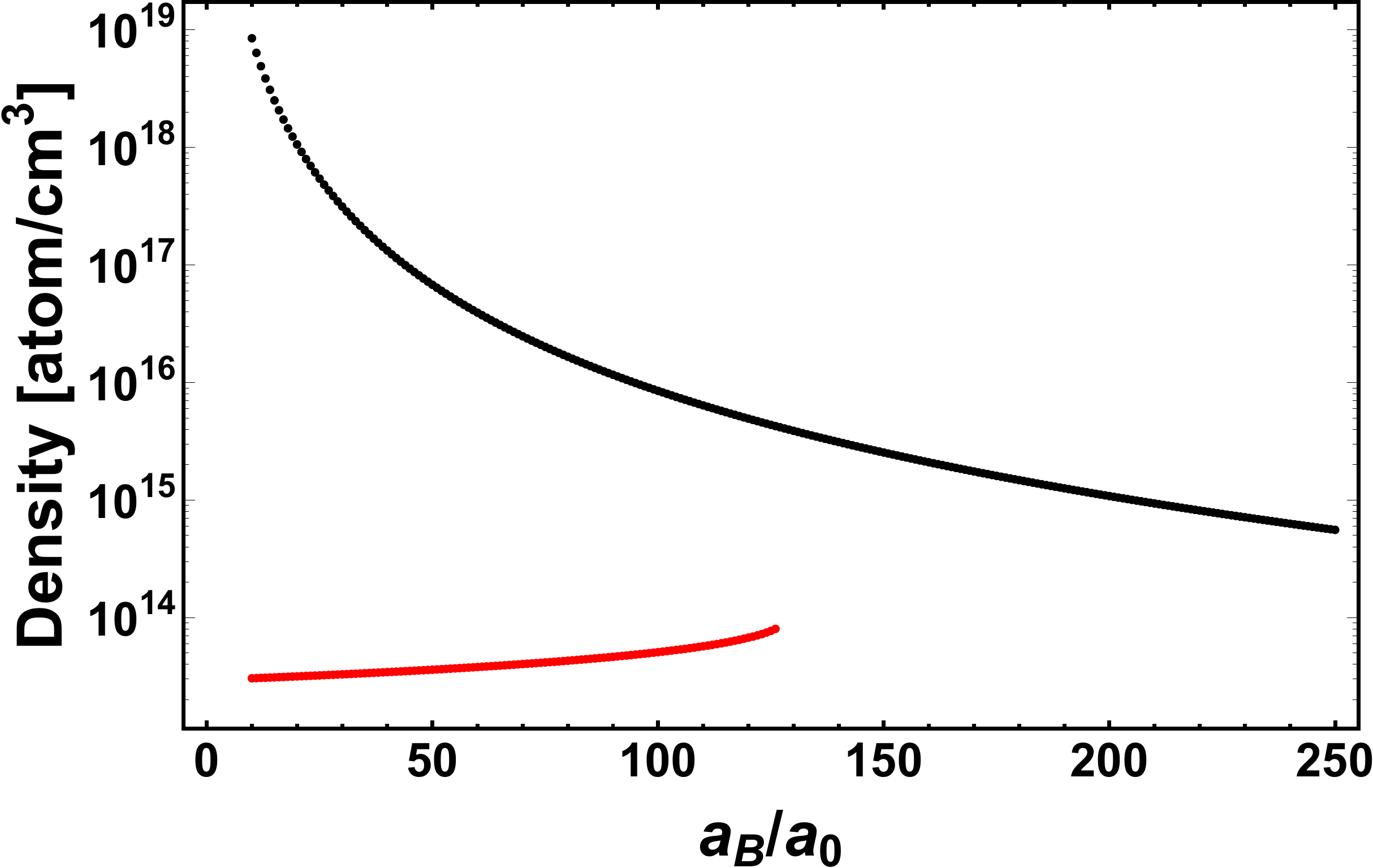} 
\caption{Energy, radial and axial widths, and a peak bosonic density of the Bose-Fermi droplet as a function of the scattering length $a_{B}$, for $a_{BF}/a_{B}=-3.6$. The numbers of bosons and fermions are $N_B=10^4$ and $N_F=620$, respectively. Black and red curves correspond to the 'droplet' and 'soliton' minima, respectively. The lighter (darker) colors for the size of the system (middle frame) represent the axial (radial) direction. } 
\label{minenergy}
\end{figure}

It turns out that the variational calculations for such parameters exhibit only one local minimum, corresponding to elongated clouds. See exemplary Fig. \ref{minenergy} confirming the existence of bifurcation in the system. For strong enough boson-fermion attraction the energy of an elongated mixture becomes negative and the trapping energy is negligibly small. It means that removing the trapping potential should not destroy the system. One should observe the oscillating Bose-Fermi droplet since its initial shape is far from being spherically symmetric. It is indeed as shown in the movies \cite{movie36,movie28}. For the first case, \cite{movie36}, we have $a_{B}=250\,a_0$ and $a_{BF}=-3.6\,a_{B}$, i.e. we are deeply in the range of parameters where the droplet exists \cite{Rakshit19a}. The trap is open in $1\,$ms. After the trap is released, the droplet survives and remains very elongated. After a further few milliseconds a rich dynamics is developed, the droplet becomes soon short in the axial and long in the radial direction. Then opposite happens and the initial shape is restored and the next cycle succeeds. Our simulations firmly prove the existence of the critical ratio $|a_{BF}|/a_{B}$ for the formation of Bose-Fermi droplets, see Ref. \cite{Rakshit19a}. The movie \cite{movie28} shows what happens with the mixture when $a_{BF}=-2.8\,a_{B}$. After a few milliseconds the droplet seems to be formed, but  eventually it explodes. For even less negative ratio, $a_{BF}/a_B=-2.0$, the mixture explodes immediately  when the trap is removed, see \cite{movie20}.

For the Bose-Fermi droplet to be formed and persist for a longer time it is necessary to minimize losses due to three-body recombination processes. For parameters as in the region marked by $1$ in Fig. \ref{LiCs}, the atomic densities are certainly too large with respect to the losses, like in Fig. \ref{minenergy}, the lowest frame. For the case of $2$ in Fig. \ref{LiCs} it seems that the creation of Bose-Fermi droplets is possible. Even more favorable conditions should be accessible for systems with higher bosonic to fermionic mass ratio (see Ref. \cite{Rakshit19a}) like, for example, bosonic ytterbium - fermionic lithium mixture. Another facilitation in creating Bose-Fermi droplets could be managed by introducing dipolar forces to the scene \cite{Schmitt16,Chomaz16}, which can effectively act as an additional, to the contact one, attraction between atoms. In such a case bosonic dysprosium/erbium - fermionic lithium mixtures could be of interest.

In summary, we study the process of formation of a Bose-Fermi mixture and predict  the bistability  in the presence of the trapping potential. For weak boson-boson repulsion (region $1$ in Fig. \ref{LiCs}), the system possesses a pair of minima, the one (related to solitons) with positive and the other (related to spherically symmetric droplets) with negative energy. For stronger intra-bosonic repulsion (region $2$ in Fig. \ref{LiCs}) only droplets related minimum survives, which represents the axially symmetric object. For strong enough boson-fermion attraction its energy becomes negative. Then, when the trapping potential is removed, the system turnes into a stable quantum Bose-Fermi droplet. It does oscillate since initially its shape is not spherically symmetric. This behavior is confirmed by quantum hydrodynamics equations based simulations. While the case $1$ in Fig. \ref{LiCs} is better suited for studying the physics of Bose-Fermi solitons, we find that the region $2$ in Fig. \ref{LiCs} could be considered as a scene for creating Bose-Fermi droplets.

\acknowledgments  
We thank C. Chin for discussions. We acknowledge support from the (Polish) National Science Center Grant No. 2017/25/B/ST2/01943. Part of the results were obtained using computers at the Computer Center of University of Bia{\l}ystok.

\appendix
\section{Variational analysis: energy contributions}
\label{energy}

Here we assume that the densities of bosonic and fermionic components are given by
\begin{eqnarray}
&&n_B(z,\rho) = \frac{N_B}{\sqrt{\pi}\, \sigma_a\, \sigma_r^2}\,\, e^{-z^2/\sigma_a^2}\,\, e^{-\rho^2/\sigma_r^2}   \nonumber \\
&&n_F(z,\rho) = \frac{N_F}{\sqrt{\pi}\, \sigma_a\, \sigma_r^2}\,\, e^{-z^2/\sigma_a^2}\,\, e^{-\rho^2/\sigma_r^2}   \,,
\label{densities}
\end{eqnarray}
hence are normalized to the number of bosons and fermions, respectively, and are of the same widths. We still admit, however, axially symmetric solutions.

All energy contributions are calculated within the local density approximation. The fermionic kinetic energy due to the Pauli exclusion principle and the gradient corrections (the Weizs{\"a}cker one) is
\begin{eqnarray}
&&E_{kin}^F = \kappa_k\, \int n_F^{5/3}\,  d^3r =  \frac{\kappa_k}{\pi} (\frac{3}{5})^{3/2} \frac{N_F^{5/3}}{\sigma_a^{2/3} \sigma_r^{4/3}}    \nonumber \\
&&E_W = \xi\, \frac{\hbar^2}{8 m_F}\, \int \frac{(\nabla n_F)^2}{n_F}\,  d^3r = \xi\, \frac{\hbar^2}{4 m_F}\, N_F (\frac{1}{\sigma_a^2} + \frac{2}{\sigma_r^2})   \,.  \nonumber \\
\label{kinfermions}
\end{eqnarray}
with $\kappa_k = (3/10)\,(6\pi^2)^{2/3}\,\hbar^2/m_F$ and $\xi=1/9$ \cite{Kirznits, Oliver}. The bosonic kinetic energy is obtained as
\begin{eqnarray}
&&E_{kin}^B = \frac{\hbar^2}{2 m_B} \int (\nabla \sqrt{n_B})^2\,  d^3r = \frac{\hbar^2}{2 m_B} N_B (\frac{1}{2 \sigma_a^2} + \frac{1}{\sigma_r^2})  \,.  \nonumber \\ 
\label{kinbosons}
\end{eqnarray}

We include contact interactions in the analysis. Within mean-field approximation the energies for boson-boson and boson-fermion interactions are
\begin{eqnarray}
&&E_{int}^B = \frac{1}{2} g_B \int n_B^2\,  d^3r =  \frac{g_B}{2 (2 \pi)^{3/2}} \frac{N_B^2}{\sigma_a \sigma_r^2}    \nonumber \\
&&E_{int}^{BF} = g_{BF} \int n_B n_F\,  d^3r = \frac{g_{BF}}{(2 \pi)^{3/2}}  \frac{N_B N_F}{\sigma_a \sigma_r^2}  \,.\nonumber \\
\label{intmf}
\end{eqnarray}
Degenerate fermions, we assume, do not interact. We consider the quantum corrections as well, including the Lee-Huang-Yang \cite{Lee57} and the Viverit-Giorgini \cite{Viverit02} ones for bosons and for bosons and fermions, respectively
\begin{eqnarray}
&&E_{LHY} = C_{LHY} \int n_B^{5/2}\,  d^3r = \frac{C_{LHY}}{\pi^{9/4}} (\frac{2}{5})^{3/2} \frac{N_B^{5/2}}{\sigma_a^{3/2} \sigma_r^3}     \nonumber \\
\label{LHY}
\end{eqnarray}
with $C_{LHY}=64/(15\sqrt{\pi})\,g_B\, a_B^{3/2}$ and
\begin{eqnarray}
&&E_q^{BF} = C_{BF}\, \int  n_B\, n_F^{4/3}\,  A(w,\alpha) \,,
\label{VG}
\end{eqnarray}
where  $w=m_B/m_F$ and $\alpha = 16\pi\, n_B a_B^3 / (6\pi^2\, n_F a_B^3)^{2/3}$ are the dimensionless parameters, and the function $A(w,\alpha)$
has a form:
\begin{eqnarray}
&& A(w,\alpha) = \frac{2(1+w)}{3w}\left(\frac{6}{\pi}\right)^{2/3}\int^{\infty}_0 {\rm d}k \int^{+1}_{-1}{\rm d}{\Omega}
\nonumber\\
&& \left[ 1 -\frac{3k^2(1+w)}{\sqrt{k^2+\alpha}}
\int^{1}_0{\rm d}q q^2 \frac{1-\Theta(1-\sqrt{q^2+k^2+2kq\Omega})}{\sqrt{k^2+\alpha}+wk+2qw\Omega}  \right]. \nonumber\\
\label{AAA}
\end{eqnarray}
The coefficient $C_{BF}=(6 \pi^2)^{2/3} \hbar^2 a_{BF}^2 / 2 m_F$.

Finally, the trapping energies are as follows
\begin{eqnarray}
E_{tr}^B &=& \frac{1}{2} m_B \int ( (\omega_a^B)^2 z^2 + (\omega_r^B)^2 \rho^2)\, n_B\, d^3r    \nonumber \\
&=&\frac{1}{2} m_B (\frac{1}{2} (\omega_a^B)^2 \sigma_a^2 + (\omega_r^B)^2  \sigma_r^2 )   \nonumber \\
E_{tr}^F &=& \frac{1}{2} m_F \int ( (\omega_a^F)^2 z^2 + (\omega_r^F)^2 \rho^2)\, n_F\, d^3r =   \nonumber \\
&=&\frac{1}{2} m_F (\frac{1}{2} (\omega_a^F)^2 \sigma_a^2 + (\omega_r^F)^2  \sigma_r^2 )  \nonumber \\
\label{trap}
\end{eqnarray}
with $\omega_a^B$ ($\omega_a^F$) and $\omega_r^B$ ($\omega_r^F$) being the axial and radial trapping frequencies for bosons (fermions).

\end{document}